\documentclass[aps,prl,showpacs,notitlepage,nofootinbib,superscriptaddress,floatfix,showkeys,twocolumn,preprintnumbers]{revtex4-1}

\usepackage{blindtext}
\usepackage{amsmath,amssymb}
\usepackage{float}
\usepackage{microtype}
\usepackage{graphicx}
\usepackage{bm}
\usepackage{latexsym}
\usepackage{epsfig}
\usepackage{psfrag}
\usepackage{color}
\usepackage[dvipsnames]{xcolor}
\usepackage{subfigure}
\usepackage[section]{placeins}
\usepackage{comment}
\usepackage{multirow}
\usepackage{enumitem}
\definecolor{PRLblue}{RGB}{0,0,150}
\definecolor{forestgreen}{HTML}{228B22}
\definecolor{tabred}{RGB}{214, 39, 40}
\definecolor{cornflowerblue}{HTML}{6495ED}
\usepackage{url}

\usepackage{tabularx}
\usepackage{booktabs} 
\usepackage{ragged2e} 

\usepackage[
  colorlinks=true,
  linkcolor=PRLblue,
  citecolor=PRLblue,
  urlcolor=PRLblue
]{hyperref}

\def\e1i{\epsilon_{1\mathrm{i}}}

\allowdisplaybreaks[1]

\makeatletter
\renewcommand\NAT@open{\textcolor{PRLblue}{[}}
\renewcommand\NAT@close{\textcolor{PRLblue}{]}}
\makeatother

\usepackage[mathlines]{lineno}

\newcommand*\patchAmsMathEnvironmentForLineno[1]{
  \expandafter\let\csname old#1\expandafter\endcsname\csname #1\endcsname
  \expandafter\let\csname oldend#1\expandafter\endcsname\csname end#1\endcsname
  \renewenvironment{#1}
     {\linenomath\csname old#1\endcsname}
     {\csname oldend#1\endcsname\endlinenomath}}
\patchAmsMathEnvironmentForLineno{equation}
\patchAmsMathEnvironmentForLineno{align}
\patchAmsMathEnvironmentForLineno{displaymath}

\usepackage{xr-hyper}
\begin{document}

\title{Decisive evidence for a global cosmic-ray feature in gamma-ray data}

\author{Fernando Valenciano}\email{fernando.valenciano@iac.es}
\affiliation{Instituto de Astrof\'{\i}sica de Canarias, C/ V\'{\i}a L\'{a}ctea, s/n, E-38205 La Laguna, Tenerife, Spain }
\affiliation{Departamento de Astrof\'{\i}sica, Universidad de La Laguna, Avenida Francisco S\'{a}nchez, s/n, E-38205 La Laguna, Tenerife, Spain}

\author{Pedro De la Torre Luque}\email{pedro.delatorre@uam.es}
\affiliation{Departamento de F\'{i}sica Te\'{o}rica, M-15, Universidad Aut\'{o}noma de Madrid, E-28049 Madrid, Spain}
\affiliation{Instituto de F\'{i}sica Te\'{o}rica UAM-CSIC, Universidad Aut\'{o}noma de Madrid, C/ Nicol\'{a}s Cabrera, 13-15, 28049 Madrid, Spain}

\author{Daniele Gaggero}\email{daniele.gaggero@pi.infn.it}
\affiliation{INFN Sezione di Pisa, Polo Fibonacci, Largo B. Pontecorvo 3, 56127 Pisa, Italy}

\author{Jorge Martin Camalich} \email{jcamalich@iac.es}
\affiliation{Instituto de Astrof\'{\i}sica de Canarias, C/ V\'{\i}a L\'{a}ctea, s/n, E-38205 La Laguna, Tenerife, Spain }
\affiliation{Departamento de Astrof\'{\i}sica, Universidad de La Laguna, Avenida Francisco S\'{a}nchez, s/n, E-38205 La Laguna, Tenerife, Spain}

\smallskip
\begin{abstract}
We analyze 17 years of \textit{Fermi}-LAT data to study the diffuse $\gamma$-ray emission along the Galactic plane and its connection to cosmic-ray (CR) propagation. 
Accounting for correlated instrumental systematics, we find decisive evidence for a spectral break at $\sim20-30$~GeV, present across the inner Galactic disk, with a preference of $\Delta\chi^2=39.78$ ($4.44 \sigma$ global). The energy and amplitude of this feature are in striking agreement with the spectral feature observed in local CR nuclei, indicating that it is a global property of the Galactic disk rather than a local phenomenon. The measured slope change, $\Delta\gamma \simeq 0.20$, is consistent with local CR data and with expectations from a transition in the CR diffusion regime from Kolmogorov to Kraichnan turbulence. Our results provide strong evidence that diffuse $\gamma$-ray emission along the Galactic disk is dominated by $\pi^0-$decay emission up to a hundred GeV. These findings provide key insights into the origin of the CR spectral hardening and the propagation of CRs throughout the Galaxy, and illustrate how this multimessenger approach can be used to probe the origin of features in the CR spectra, such as the long-studied CR Knee.
\end{abstract}
\maketitle

\textbf{A multimessenger approach to understand the origin of CR spectral features ---}
The last two decades have witnessed a major advancement in precision and energy coverage of both charged Galactic cosmic-ray (CR) and $\gamma$-ray measurements~\cite{Adriani2013PAMELA, An2019DAMPE, PhysRevLett.114.171103_AMSprotons, Yoon2017CREAMIII, Kobayashi2022CALETp, Abeysekara2019HAWC, Cao2023LHAASODiffuse, Amenomori2021DiffusePeV}. 
Traditionally, these data were interpreted using a phenomenological framework (recently reviewed in \cite{Gabici:2019jvz}) that typically featured a ``universal'' injection spectrum of sources and a featureless power-law scaling of the diffusion coefficient with rigidity up to the CR knee.
\\
\indent However, several anomalies have been detected over the years in different channels, and the conventional scenarios describing CR acceleration and propagation are under pressure. 
A key development was the discovery of breaks in the spectra of protons, Helium, light and heavy nuclear particles at rigidities $R \simeq 300$ GV, not clearly observed by PAMELA \citep{Adriani_2011} and detected with high precision by AMS-02 \citep{PhysRevLett.114.171103_AMSprotons, PhysRevLett.123.181102_AMSHe}. Below the break, the proton spectrum is slightly softer with respect to the other nuclei ($\gamma \simeq 2.85$ for the former, $\gamma \simeq 2.75$ for the heavier nuclei), while above the break, all spectra progressively harden to values around $\gamma \simeq 2.6$. 
\citet{Genolini:2017dfb} have analyzed the Boron/Carbon data claiming {\it decisive evidence} in favor of a propagation origin of this spectral feature. 
\\
\indent However, the origin of this feature remains an open question. Non-linear effects and inhomogeneous diffusion may play a role as far as the physical interpretation is concerned~\cite{Gabici:2019jvz, Aloisio_2013, Ptuskin_2013}. For instance, \citet{Blasi:2012yr} proposed to connect this feature to a transition from a low-energy regime where CR particles are confined by self-generated turbulence to a high-energy regime where CRs are scattered by pre-existing turbulence. \citet{Tomassetti:2012ga} presented instead a phenomenological model featuring a different rigidity dependence of diffusion in the low-latitude region close to the Galactic disk with respect to the diffusive halo. 
\begin{figure}[!t]
    \centering
    \includegraphics[width=0.95\linewidth]{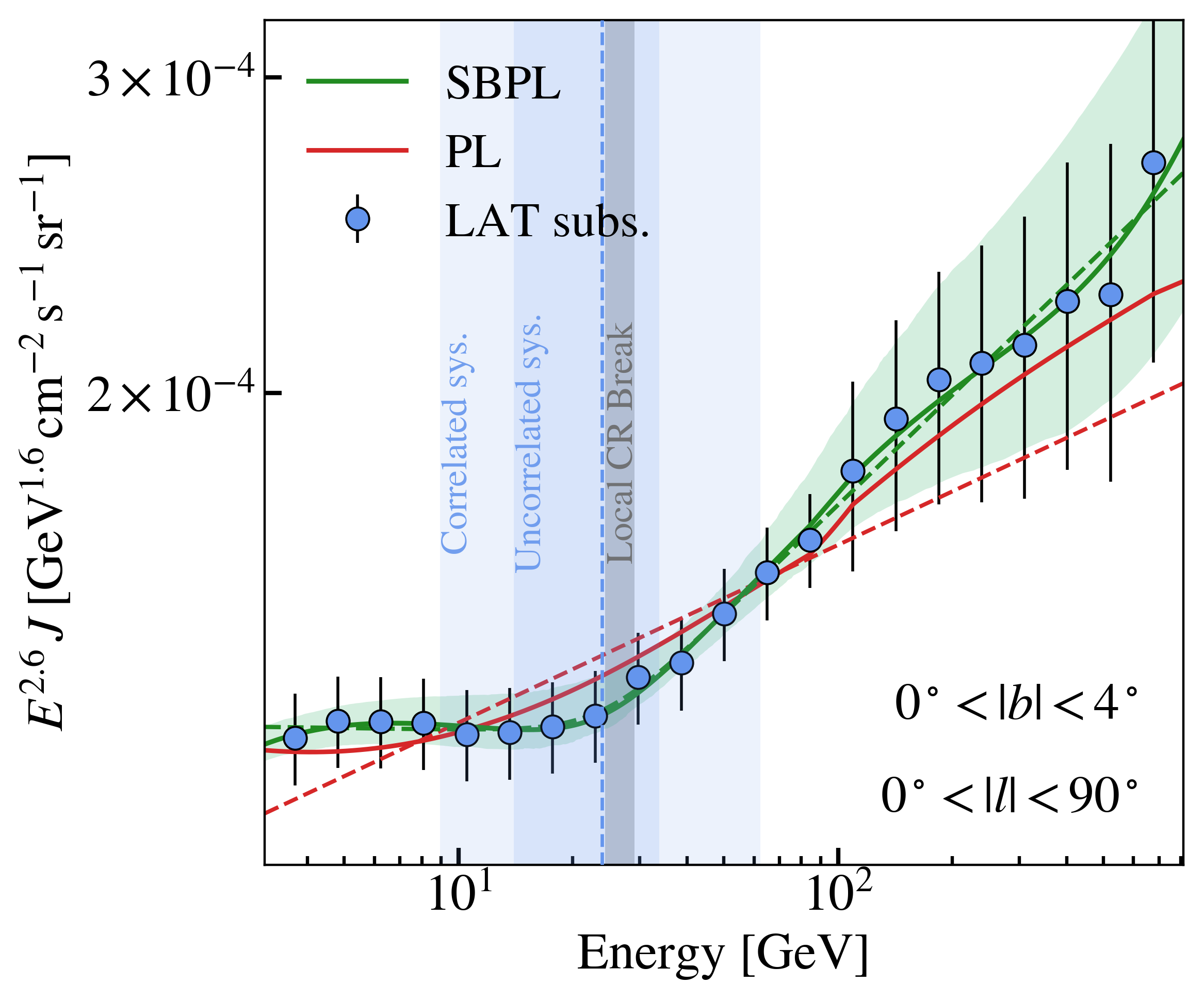}
    \caption{\textit{Fermi}-LAT spectral energy distribution (SED) for $|b|<4^\circ$, averaged over $0<|l|<90^\circ$, after subtraction of the foreground components. Error bars combine statistical and effective-area systematic uncertainties in quadrature. The spectral break is clearly visible in the residual emission. The unbroken (PL; red), and smoothly broken (SBPL; green) power-law are shown, together with the 1$\sigma$ uncertainty band for the SBPL case. Dashed lines are the standard fits; solid lines are the fits obtained when the energy-correlated effective-area nuisance parameters are left free and set to their best-fit values. For completeness, we also show the energy band uncertainty without considering correlations.
    }
    \label{fig:fig1}
\end{figure}

Regardless of the origin, this spectral feature should imprint itself in the diffuse $\gamma$-ray emission, as hadronic interactions of CRs with the interstellar medium produce $\gamma$ rays with energies $E_\gamma \sim 0.09 E_p$ \citep{Berezinsky1990, Aharonian2004}. Consequently, observations of the diffuse $\gamma$-ray emission could provide a powerful probe of the feature's origin~\cite{Dermer_2015, Fermi-LAT:2016zaq, 2014PhRvL.112o1103A}.
In the simplest case, a uniform break at $R \sim 300$ GV in protons would translate into a spectral feature at $E_\gamma \sim 20$--30 GeV in the diffuse $\gamma$-ray emission. Detecting such a feature in \textit{Fermi}-LAT data would open the possibility of studying its origin through $\gamma$-ray observations. In particular, mapping its spatial dependence across the Galaxy would provide a novel test of inhomogeneous CR transport~\citep{Tovar-Pardo:2024cbp}, while distinguishing this scenario from alternative explanations. Remarkably, a method to statistically assess the global significance and spatial variation of a spectral feature in any non-thermal Galactic diffuse emission represents a powerful diagnostic to characterize the origin of generic CR spectral features, which typically represent the key tell-tale signs of the most crucial physical aspects of particle acceleration and confinement.

In this work, we exploit more than 17 years of \textit{Fermi}-LAT observations and present the first rigorous characterisation of a spectral break in the diffuse Galactic $\gamma$-ray emission, as shown in Fig.~\ref{fig:fig1}.
We further find that this break appears consistently across all longitude windows along the Galactic plane, indicating that this feature is global and mirrors the behaviour observed in local CR data. 

\textbf{ Analysis of the $\gamma$-ray diffuse emission ---}
We analyze \textit{Fermi}-LAT diffuse emission along the Galactic plane within latitudes $|b|<4^\circ$ and over the energy range $2$--$1000$~GeV, where hadronic emission dominates and the spectral break is expected to manifest. To probe the spatial universality, we divide the Galactic plane into longitude windows of width $\Delta l=10^\circ$ (with a central bin $|l|<5^\circ$) restricted to $|l|<85^\circ$, where sufficient photon statistics are available for a meaningful spectral fit.
This binning is chosen to balance spatial resolution against photon statistics: narrower windows would lack sufficient high-energy events for a meaningful spectral fit, while wider windows would dilute spatial information. We fit only energy bins containing at least 10 photon counts, allowing each window to retain sufficient photons up to nearly 800 GeV. \\

\emph{Fermi-LAT diffuse data ---}
We employ 17 years of data (206 months, from 4 August 2008 to 4 October 2025), selecting CLEAN events from the PASS8 dataset with the P8R3\_CLEAN\_V3 instrument response functions. Standard data quality cuts were applied throughout, including \texttt{(DATA\_QUAL>0) \&\& (LAT\_CONFIG==1)}, 
rocking angles $\theta_r \leq 52^\circ$ and a zenith-angle cut of $\theta_z < 90^\circ$ to suppress contamination from Earth albedo emission.
Both front- and back-converted events were retained to maximize photon statistics.
Data extraction and exposure map generation were performed using the most recent release of the ScienceTools \footnote{\url{https://fermi.gsfc.nasa.gov/ssc/data/analysis/software/};  \url{https://github.com/fermi-lat/Fermitools-conda/wiki/Installation-Instructions}}, spanning an energy range from $30$ MeV up to $1$~TeV divided into 40 logarithmically spaced bins.
\\

\emph{Subtraction of foreground components --}
Since we are primarily interested in the diffuse hadronic emission — dominated by CR nuclei interacting with the interstellar gas — we evaluate and subtract the principal non-hadronic contributions to the total emission (see Supplemental Material (SM), Sec. I for details). These are:
\begin{enumerate}[label=\roman*), itemsep=0pt, leftmargin=6mm, nosep]
    \item \textit{Resolved point-sources}: All-sky maps are constructed by integrating the best-fit spectral model of each source in the 4FGL-DR4 catalog~\cite{2023arXiv230712546B} over the considered energy range, and spatially convolved with the energy-dependent LAT point-spread function (PSF)\footnote{\url{https://fermi.gsfc.nasa.gov/ssc/data/analysis/documentation/Cicerone/Cicerone_LAT_IRFs/IRF_PSF.html}}.
    \item \textit{Isotropic background}: Modelled from the official \textit{Fermi}-LAT template (iso\_P8R3\_CLEAN\_V3\_v1)\footnote{\url{https://fermi.gsfc.nasa.gov/ssc/data/access/lat/BackgroundModels.html}}, which includes extragalactic emission and residual detector background.
    \item \textit{Fermi bubbles}: We used the publicly available spatial and spectral templates derived in Ref.~\citep{2017ApJ...840...43A}\footnote{\url{https://www-glast.stanford.edu/pub\_data/1220/}}. 
    \item \textit{Inverse-Compton (IC)}: Arising from the interaction of CR electrons with interstellar radiation fields, computed following Ref.~\cite{DeLaTorreLuque:2025zsv}, where the steady-state lepton distribution is tuned to local CR electron and positron data.
    \item \textit{Bremsstrahlung}: From the interaction of CR electrons with interstellar gas, evaluated following Ref.~\cite{DeLaTorreLuque:2025zsv}.
    \item \textit{Unresolved sources:} Sources below the \textit{Fermi}-LAT detection threshold that cannot be individually disentangled from the truly diffuse background are estimated following Refs.~\cite{Luque:2022buq, DeLaTorreLuque:2025zsv} derived in Ref.~\cite{Steppa2020}.
\end{enumerate} 
Among these components, the resolved point-source is the most significant contribution. Above a few GeV, bremsstrahlung is negligible, and IC emission remains subdominant in the plane at the $\sim$10\% level. The isotropic background and Fermi bubble contributions are minimal at low Galactic latitudes, and the unresolved-source contribution remains subdominant across the \textit{Fermi}-LAT energy range. 
Overall, we assess the impact of each component by subtracting them individually and in combination, finding no significant variation in the diffuse emission spectra, consistent with the Galactic plane emission being dominated by hadronic interactions in the energy range considered.
\\

\textit{Spectral analysis ---} The hadronic diffuse emission spectrum in each longitude window is fitted independently under two nested hypotheses: a single power-law (PL), representing the null hypothesis of no spectral break, 
\begin{equation}
    \mu^{\rm PL}(E)=K\left(\frac{E}{E_0}\right)^{-\gamma}
\end{equation}
where $K$ is the normalisation and $\gamma$ the slope, and a smoothly broken power-law (SBPL), parameterizing a gradual hardening around a characteristic break energy $E_{\rm break}$ with slope change $\Delta\gamma$
\begin{equation}\label{eq:sbpl}
    \mu^{\rm SBPL}(E)=K\left(\frac{E}{E_0}\right)^{-\gamma}\left[1+\left(\frac{E}{E_{\rm break}}\right)^{\Delta\gamma/s}\right]^{-s}
\end{equation}
and $s$ measures the smoothness of the transition, we adopt a fixed value of $s=0.04$ motivated by local CR data~\citep{Genolini:2017dfb, Genolini:2019ewc} -- see SM, Sec. II for the corresponding spectral fits. The statistical preference for a break over the null hypothesis is quantified via a $\chi^2$ goodness-of-fit test, combining statistical Poissonian and systematic effective-area uncertainties in quadrature \footnote{Summing them linearly does not affect our results significantly, since systematic errors largely dominate.}. To rigorously assess whether the observed break could arise from energy-correlated instrumental systematics rather than a physical spectral feature, we introduce smooth nuisance deformations of the predicted spectrum, constrained by uniform priors and profiled independently in each longitude window --- following the statistical approach of a previous \textit{Fermi}-LAT analysis~\citep{2017PhRvD..95h2007A} (see SM, Sec. III for details). 
The preference for a break is quantified through the profiled test statistic $\Delta\chi^2_{\rm prof}=\chi^2_{\rm PL,prof}-\chi^2_{\rm SBPL, prof}$, evaluated both per longitude window and in a \textit{global} fit in which the break parameters are shared across all windows.  
Statistical significance is then quoted in two ways. First, $\Delta\chi^2$ is converted to a $p-$value assuming Wilks' theorem~\citep{10.1214/aoms/1177732360}, yielding $Z_{\rm Wilks}$. Second, a Monte Carlo (MC) approach is employed: $2\times10^6$ realizations of the null hypothesis -- PL spectra with Poisson noise and energy-correlated effective area uncertainties -- are generated and analyzed through the same fitting procedure. The fraction of those whose $\Delta\chi^2_{{\rm prof},i}$ exceeds the observed value is the empirical $p-$value, then converted to $Z_{\rm MC}$. This provides a reliable inference under the non-regular statistical conditions arising from bounded nuisance parameterizations.
\\

\begin{figure}[t!]
    \centering
    \includegraphics[width=\linewidth]{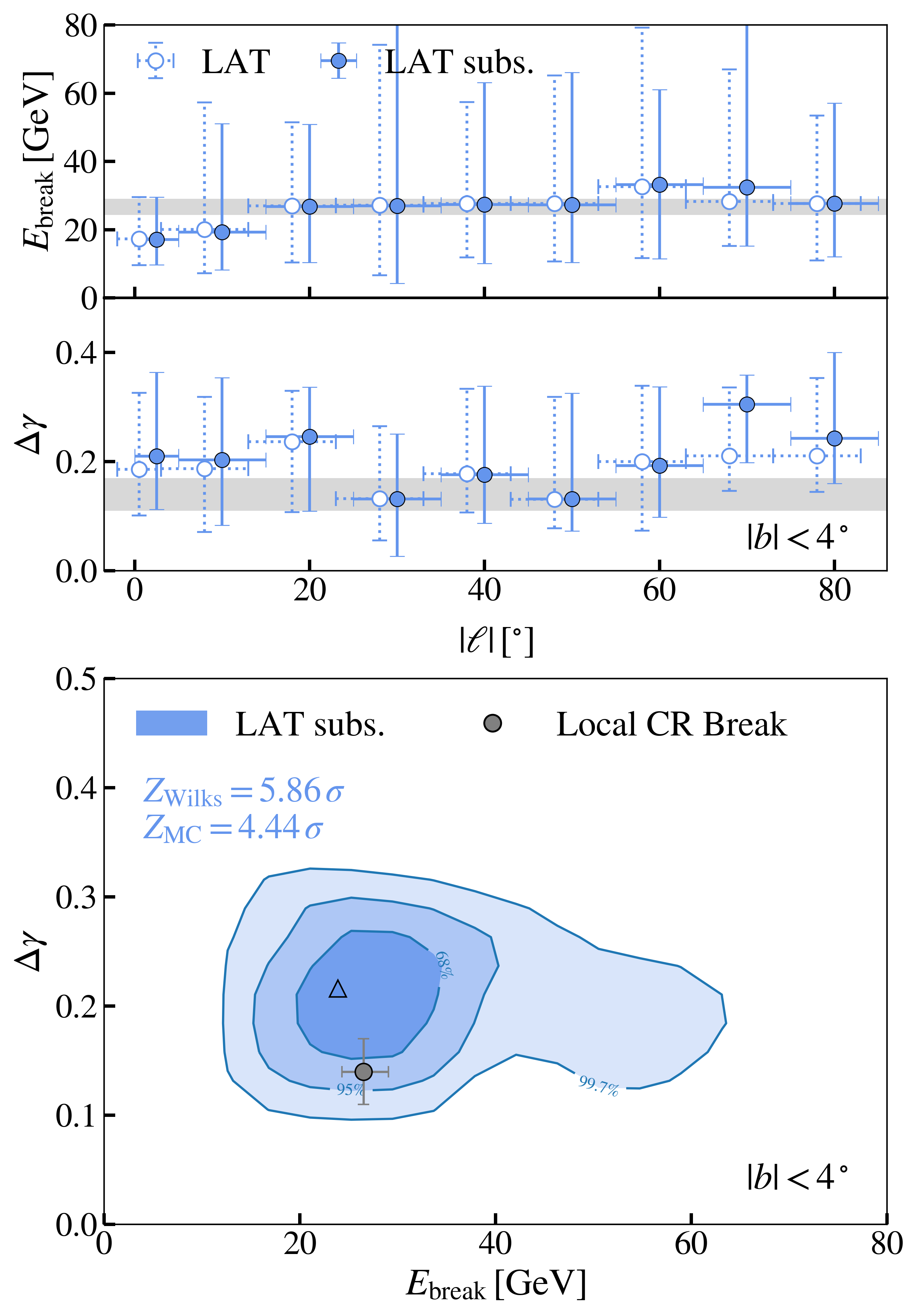}
    \caption{\textit{Top panel:} Best-fit break energy $E_{\mathrm{break}}$ (top) and spectral-index change $\Delta\gamma$ (bottom) from SBPL parameterisation (Eq.~\ref{eq:sbpl}) inferred from \textit{Fermi}-LAT data. The gray band indicates the expectation from local CR measurements, assuming $E_\gamma \simeq 0.09\,E_p$. \textit{Bottom panel:} Global-fit confidence contours for a model with a common $E_{\mathrm{br}}$ and $\Delta\gamma$ across nine longitude windows. Confidence levels are constructed using Wilks' theorem from the profiled statistic $\Delta\chi^2_{\mathrm{prof}}$ and are shown for comparison with the local CR break. $Z_{\rm Wilks}$ 
    and $Z_{\rm MC}$ denote the significance estimated via Wilks' theorem 
    and MC realizations, respectively.}
    \label{fig:ebreak}
\end{figure}

\textbf{Results: The CR rigidity break across the Galactic Plane ---}
Fig.~\ref{fig:fig1} shows the spectral energy distribution of the diffuse hadronic emission for $|b|<4^\circ$, summed over the full $0^\circ<|l|<85^\circ$ Galactic disk analysed. A clear departure from the unbroken PL fit emerges above $\sim20-30$ GeV, well captured by the SBPL model; the shaded bands show the effect of correlated and uncorrelated effective-area systematics in the determination of the break position.
Fig.~\ref{fig:ebreak} summarizes the principal outcome of our analysis. The inferred break energies cluster consistently around $E_{\rm break}\sim 20-30$ GeV across all longitude windows, in excellent agreement with the expectations from a rigidity feature at $R\sim300$ GV in CR protons, through the relation $E_\gamma\sim0.09\,E_p$~\citep{Berezinsky1990, Aharonian2004}. The measured slope change, $\Delta\gamma\sim0.20$, is remarkably consistent with values inferred from CR nuclei analysis~\citep{Genolini:2017dfb, Genolini:2019ewc}.

We evaluate $\Delta\chi^2_{\rm prof}$ in three complementary configurations: window-by-window independently; a \textit{global} fit with break parameters shared across all longitude windows; and a global fit with break parameters fixed to those inferred from local CR rigidity analyses (\textit{fixed}). In the window-by-window analysis, the MC significance is individually modest, with an average $\langle Z_{\rm MC}\rangle\simeq 1.52\sigma$, reflecting the limited statistical power of each individual window. When fitting all windows simultaneously with shared break parameters in the global fit, precision and significance increase substantially since this approach mitigates regional smoothing and enhances the signal-to-noise ratio imposing a more restrictive null hypothesis -- a PL spectrum must coherently reproduce a break-like pattern across all the independent windows. The full results are summarized in Table~\ref{tab:summary}.

\begin{table}[t]
\vspace{1mm}
\centering
\caption{Summary of fit quality and spectral break parameters for the global analysis with shared break parameters across longitude windows (\textit{global}) and with fixed shared break parameters to CR measurements (\textit{fixed}).}
\label{tab:summary}
\renewcommand{\arraystretch}{1.4} 
\setlength{\tabcolsep}{4pt}
\begin{tabular*}{\columnwidth}{@{\extracolsep{\fill}}lccccc}
\hline\hline
\textbf{SBPL} & $\Delta\chi^2$ & $E_{\rm break}$ [GeV] & $\Delta\gamma$ & $Z_{\rm W}$ [$\sigma$] & $Z_{\rm MC}$ [$\sigma$]\\
\hline
Global & 39.78 & $24.08^{+9.06}_{-1.63}$  & $0.20^{+0.05}_{-0.02}$ & 5.86 & $4.44$  \\
Fixed  & 35.49 & 26.52      & 0.14     & ---  &  $4.28$  \\
\hline\hline
\end{tabular*}
\end{table}

Even under the most conservative treatment of correlated systematic uncertainties 
the global fit yields $\Delta\chi^2_{\rm prof}=39.78$ with $Z_{\rm Wilks}=5.86\sigma$ and $Z_{\rm MC}=4.44\sigma$ (with only nine out of $2\times10^6$ MC realizations under the null hypothesis exceeding the SBPL model).
Indeed, when the break parameters are fixed to the values inferred from local CR observations, the two hypotheses contain the same number of free parameters and the likelihood ratio can be directly interpreted in terms of a Bayes factor. We obtain $\Delta\chi^2 = 2\ln\kappa = 35.49$, which corresponds to {\it decisive} evidence on the Jeffreys scale~\cite{Jeffreys:1939xee}, strongly favoring the interpretation that the observed $\gamma$-ray feature coincides with the spectral feature seen in local CRs.
Moreover, the joint constraints in the $(E_{\rm break},\Delta\gamma)$ plane, shown in Fig.~\ref{fig:ebreak} (\textit{bottom}), demonstrate a clear and consistent preference for a broken spectrum, in compelling agreement with expected values from local CR measurements.
\\

\textit{Robustness tests.}—
The preference for a spectral break and the reconstructed $(E_{\rm break},\Delta\gamma)$ parameters remain stable under a wide range of cross-checks, which are detailed in the SM, Sec. IV:
\begin{enumerate}[label=\roman*), itemsep=0pt, leftmargin=5mm, nosep]
\item \textit{Analysis regions:} Adopting a tighter latitude cut of $|b|<2^\circ$ yields consistent results with moderately larger uncertainties due to reduced photon statistics, and the break persists across all alternative window divisions explored, from coarser four-window to finer eighteen-window divisions of the Galactic plane. 

\item \textit{Foreground components:} Since the point-source contribution is the largest, we have verified that doubling its normalisation leaves the spectral break parameters unchanged, confirming that the feature is not driven by source contamination. Subtracting the full set of non-hadronic diffuse components -- isotropic emission, Fermi bubbles, IC and bremsstrahlung templates, and unresolved sources -- leaves the break parameters essentially unmodified, as shown in Fig.~\ref{fig:ebreak} (\textit{top}), where the unsubtracted and foreground-subtracted results are in excellent  agreement across all longitude windows, confirming that the feature is not induced by any of the subtracted components.
\item \textit{IEM consistency check:} The official \textit{Fermi}-LAT Interstellar Emission Model (IEM) is a full-sky template of the diffuse $\gamma$-ray emission, encoding contributions from hadronic, IC, and bremsstrahlung emission through the GALPROP CR propagation framework tuned to the LAT data~\cite{Fermi-LAT:2016zaq}. Since it provides an independent characterisation of the diffuse emission that does not rely on our point-source subtraction procedure, it serves as a powerful consistency check of our results. Processing this template through an identical analysis pipeline yields break parameters fully consistent with those derived directly from \textit{Fermi}-LAT data, confirming that the observed feature is not an artifact of our specific foreground prescription.
\item \textit{Nuisance configurations:} Testing smooth and non-smooth deformation functions with box-constrained and Gaussian priors yields nearly identical break parameters throughout.
\end{enumerate}
Overall, these tests establish the robustness of the detection against any plausible systematic effects and analysis choices, providing compelling evidence for a common spectral break in the diffuse Galactic $\gamma$-ray emission along the plane. The precise agreement with the well-established rigidity break observed in local CR nuclei offers a largely model-independent test to further understand the origin of this break and an indication that the $\sim$300~GV break is a global property of CR transport throughout the Galactic plane, rather than a local peculiarity of the Solar neighborhood.\\

\textbf{Implications and conclusions ---}
Our study exploits $\gamma$ rays as a probe of Galactic CR propagation. Since $\gamma$ rays are predominantly produced by CR nuclei interactions with the interstellar medium, their diffuse emission encodes information about CR densities, spectra, and transport processes across the Galactic disk. By examining this emission with high precision, we connect observed $\gamma$-ray features directly to properties of the underlying CR population, providing an independent test of CR propagation and turbulence regimes across the Galaxy.

We present the first statistically rigorous and spatially resolved characterisation of a spectral break in the diffuse Galactic $\gamma$-ray emission appearing uniformly across the Galactic plane, with a break energy $E_{\rm break}\sim 20-30$~GeV and a slope change $\Delta\gamma\sim0.20$, in striking agreement with the $\sim300$~GV rigidity break observed in local CR nuclei.
Even under the most conservative treatment of correlated systematic uncertainties, the global significance of this feature reaches $Z_{\rm Wilks}=5.86\sigma$, with Monte Carlo tests confirming $Z_{\rm MC}=4.44\sigma$.
The break is stable against variations in window size, latitude cuts, foreground prescription and systematic treatment of uncertainties, confirming it is not an artifact of the analysis.

A key and unexpected finding is the remarkable spatial uniformity of the break across the Galactic plane: both the break energy and the change in slope are consistent with a single common value across all longitude windows. This spatial coherence may provide important clues to the origin of the $\sim300$~GV feature. In particular, it is difficult to reconcile with scenarios in which the break emerges from the superposition of different source injection spectra~\cite{2012MNRAS.421.1209T, Liu:2018fjy}, since variations in the source population would generically lead to spatial variations in the position and/or shape of the feature. Instead, a transport-related origin appears more compatible with the observed coherence, although the result is itself non-trivial for such scenarios. For instance, if the break marks the transition between confinement by self-generated and pre-existing turbulence, its characteristic energy depends on several physical quantities that are in principle expected to vary across the Galaxy: the observed uniformity suggests that these variations combine in such a way as to leave the transition energy nearly unchanged.

The hadronic origin of the feature is strongly supported by several arguments. Any IC contribution is subdominant in the Galactic plane where $\pi^0$ decay dominates above a few GeV. The break energy coincides with the hardening in local CR proton spectra, while the local electron spectrum shows no corresponding hardening at these energies. Furthermore, an IC-generated break would become more prominent with increasing latitudes as the $\pi^0$/IC ratio decreases -- contrary to what we observe. 
The observed break parameters and their spatial uniformity are therefore not naturally accommodated by plausible IC models.

In a broader perspective, this work can pave the way towards a research program centered on highlighting the imprint of CR transport effect on non-thermal radiation. In fact, a comprehensive remapping of charged CRs spectral breaks to corresponding spectral features in the radio/$\gamma$-ray or even neutrino emission at different energies will provide a unique tool to diagnose the origin of such breaks and reveal the underlying physics. A useful application of these methods can be the study of the $10$~TV structure observed by DAMPE in the proton and helium spectra, which should leave a corresponding imprint in $\gamma$ rays around $1$~TeV, within reach of CTA and SWGO. At even larger energies, a further application is the study of the PeV domain, where an inconsistency between the position of the local CR knee as measured by LHAASO, and the associated diffuse $\gamma$-ray emission detected by the same experiment appears in mild tension \cite{Castro:2025wgf}.

\vspace{0.7cm}

\noindent\begin{minipage}{\linewidth}
\footnotesize
\setlength{\baselineskip}{0.9\baselineskip} 
\emph{Acknowledgements---} 
We thank Mario Giliberti for producing some of the datasets used in this analysis, and Philippe Bruel for the help provided with the treatment of the instrumental uncertainties.
FV acknowledges support from the European Union through the grant ``UNDARK'' of the Widening participation and spreading excellence programme (project number 101159929). 
FV also acknowledges the MICINN through the grant ``DarkMaps'' PID2022-142142NB-I00.
PDL has been supported by the Juan de la Cierva JDC2022-048916-I grant, funded by MCIU/AEI/10.13039/501100011033 European Union "NextGenerationEU"/PRTR, and is currently supported by Ramón y Cajal RYC2024-048445-I grant, which is funded by MCIU/AEI/10.13039/501100011033 and FSE+. The work of PDL is also supported by the grants PID2021-125331NB-I00 and CEX2020-001007-S, both funded by MCIN/AEI/10.13039/501100011033 and by ``ERDF A way of making Europe''. PDL also acknowledges the MultiDark Network, ref. RED2022-134411-T. 
DG acknowledges support from the  Research grants TAsP (Theoretical Astroparticle Physics) and TEONGRAV funded by \textsc{infn}.
This project used computing resources from the National Academic Infrastructure for Supercomputing in Sweden (NAISS) under project NAISS 2024/5-666.
The \textit{Fermi}-LAT Collaboration acknowledges generous ongoing support from a number of agencies and institutes that have supported both the development and the operation of the LAT as well as
scientific data analysis. These include the National Aeronautics and Space Administration and the
Department of Energy in the United States, the Commissariat à l’Energie Atomique and the Centre
National de la Recherche Scientifique / Institut National de Physique Nucléaire et de Physique des
Particules in France, the Agenzia Spaziale Italiana and the Istituto Nazionale di Fisica Nucleare in
Italy, the Ministry of Education, Culture, Sports, Science and Technology (MEXT), High Energy Accelerator Research Organization (KEK) and Japan Aerospace Exploration Agency (JAXA) in Japan,
and the K. A. Wallenberg Foundation, the Swedish Research Council and the Swedish National Space
Board in Sweden. Additional support for science analysis during the operations phase is gratefully
acknowledged from the Istituto Nazionale di Astrofisica in Italy and the Centre National d’Etudes
Spatiales in France. This work performed in part under DOE Contract DE- AC02-76SF00515.
The authors would like to thank Jean-Marc Casandjian and Melissa Pesce-Rollins for their
helpful comments and suggestions during the internal review process.

\end{minipage}
\vspace{0.6em}

\bibliographystyle{apsrev4-1}
\bibliography{CRbiblio}

\end{document}


\title{Supplemental Material:\\ Decisive evidence for a global cosmic-ray feature in gamma-ray data}

\author{Fernando Valenciano}\email{fernando.valenciano@iac.es}
\affiliation{Instituto de Astrof\'{\i}sica de Canarias, C/ V\'{\i}a L\'{a}ctea, s/n, E-38205 La Laguna, Tenerife, Spain }
\affiliation{Departamento de Astrof\'{\i}sica, Universidad de La Laguna, Avenida Francisco S\'{a}nchez, s/n, E-38205 La Laguna, Tenerife, Spain}

\author{Pedro De la Torre Luque}\email{pedro.delatorre@uam.es}
\affiliation{Departamento de F\'{i}sica Te\'{o}rica, M-15, Universidad Aut\'{o}noma de Madrid, E-28049 Madrid, Spain}
\affiliation{Instituto de F\'{i}sica Te\'{o}rica UAM-CSIC, Universidad Aut\'{o}noma de Madrid, C/ Nicol\'{a}s Cabrera, 13-15, 28049 Madrid, Spain}

\author{Daniele Gaggero}\email{daniele.gaggero@pi.infn.it}
\affiliation{INFN Sezione di Pisa, Polo Fibonacci, Largo B. Pontecorvo 3, 56127 Pisa, Italy}

\author{Jorge Martin Camalich} \email{jcamalich@iac.es}
\affiliation{Instituto de Astrof\'{\i}sica de Canarias, C/ V\'{\i}a L\'{a}ctea, s/n, E-38205 La Laguna, Tenerife, Spain }
\affiliation{Departamento de Astrof\'{\i}sica, Universidad de La Laguna, Avenida Francisco S\'{a}nchez, s/n, E-38205 La Laguna, Tenerife, Spain}

\maketitle

\section{Foreground components}
\label{sm:contributions}

\begin{figure}[!hbp]
    \centering
    \begin{minipage}[t]{0.46\linewidth}
        \vspace{0pt} 
        \includegraphics[width=\linewidth]{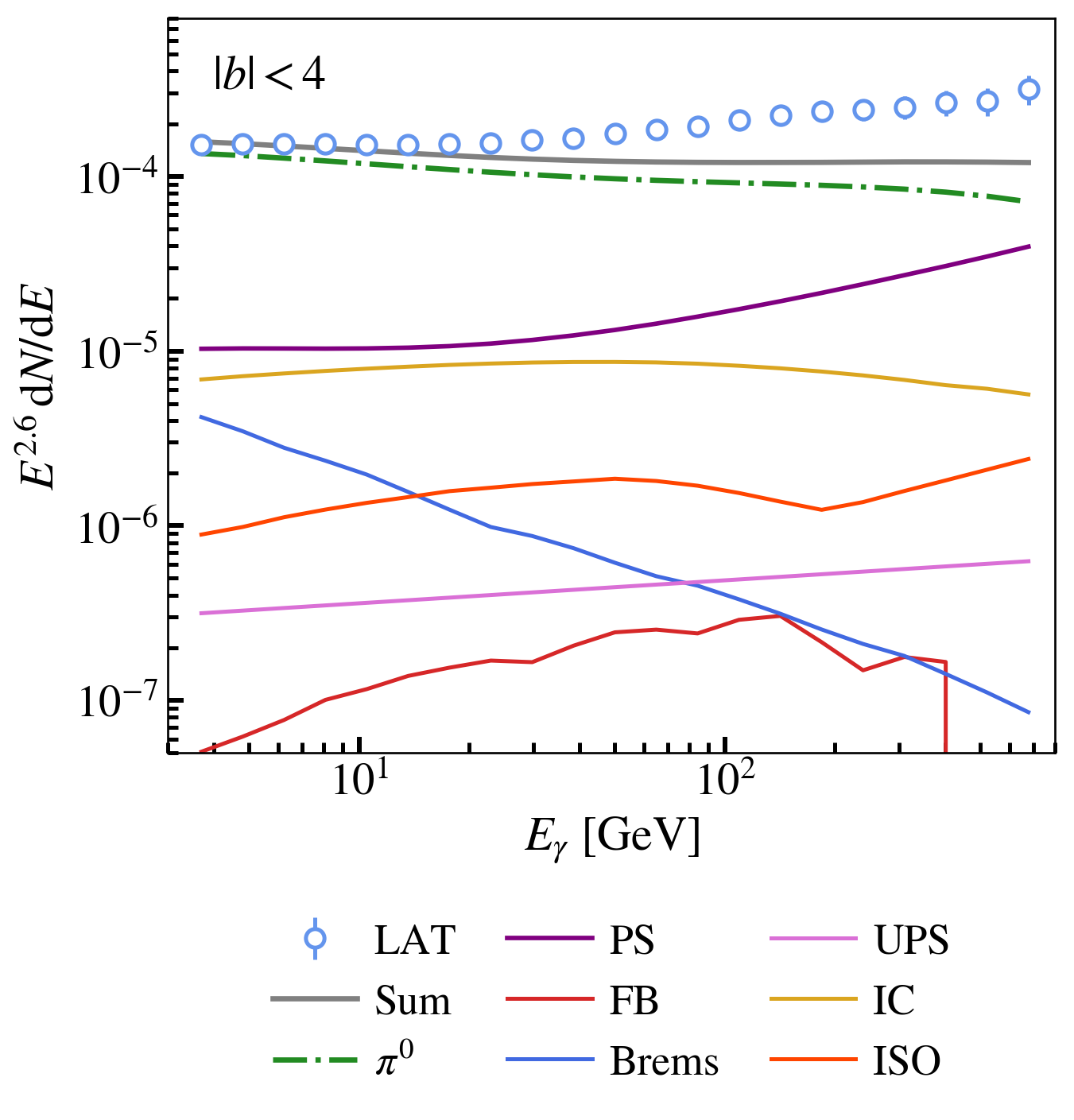}
    \end{minipage}
    \hfill
    \begin{minipage}[t]{0.53\linewidth}
        \vspace{0pt}
        \includegraphics[width=\linewidth]{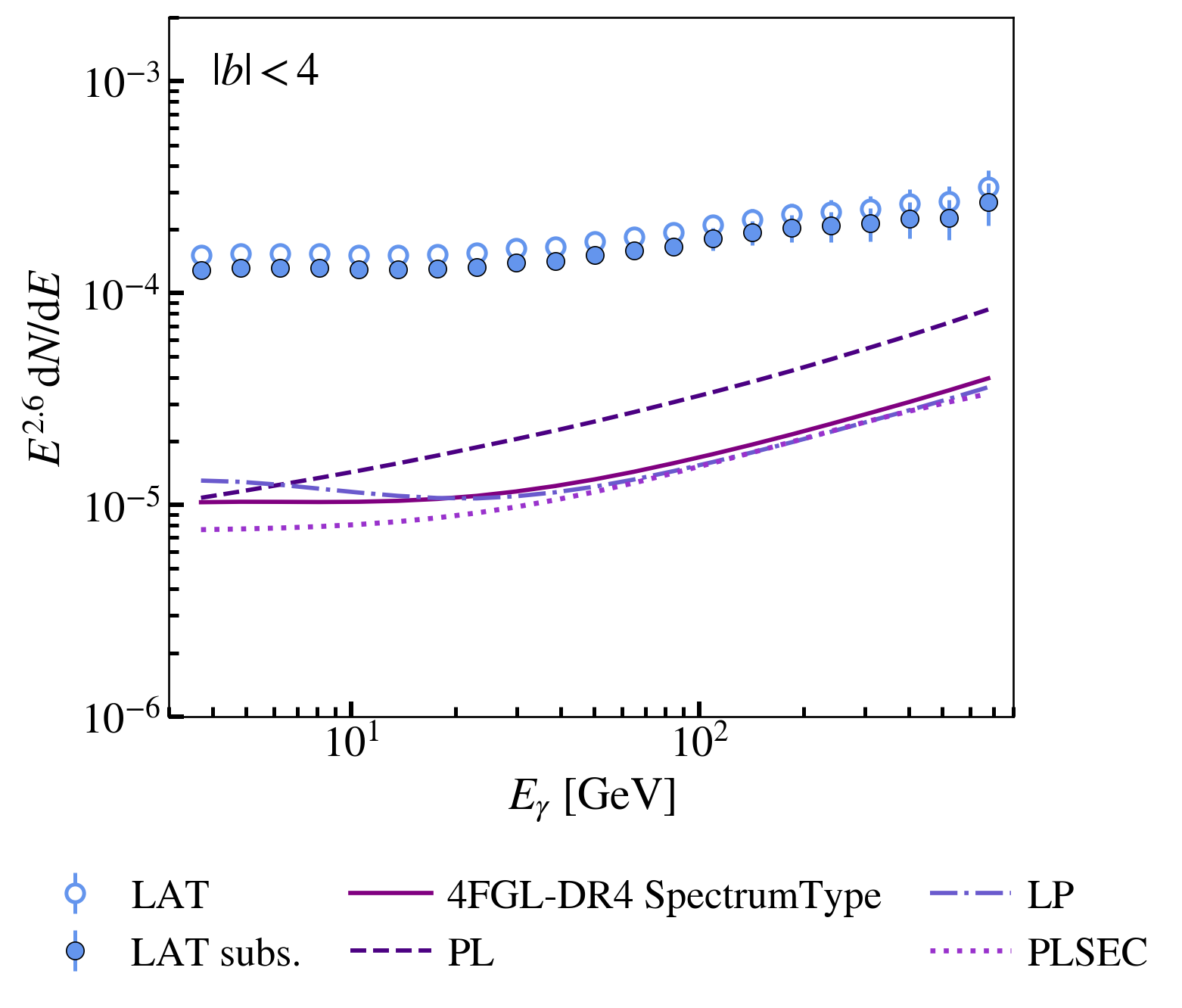}
    \end{minipage}
     \caption{\textit{Left:} \textit{Fermi}-LAT data spectral energy distribution (SED) for $|b|<4^\circ$, averaged over $0^\circ<|l|<90^\circ$, shown together with the individual foreground components. The spectral break persists regardless of the foreground-subtraction scheme adopted, showing that the diffuse $\gamma$-ray emission in the plane is dominated by hadronic emission. Solid lines show the individual foreground components used in the subtraction; for visual comparison only, we additionally show the expected $\pi^0$-decay emission from DRAGON assuming a power-law injection spectrum, which is not itself subtracted.
    \textit{Right:} Comparison of point-source emission modeled with different spectral parameterizations (power law, log-parabola, and power law with super-exponential cutoff). The solid line corresponds to the total emission obtained using the best-fit spectral model for each source in the 4FGL-DR4 catalog.}
    \label{fig:app_background}
\end{figure}

In this section, we provide technical details for the  foreground components used to isolate the hadronic diffuse emission in \textit{Fermi}-LAT data, shown in Fig. \ref{fig:app_background} (\textit{left}).

i) \textit{Resolved point-sources ---}
\label{app:ps_template}
We construct the point-source template using best-fit spectral models from the 4FGL-DR4 \textit{Fermi}-LAT catalog \cite{2023arXiv230712546B}, namely a power-law (PL), a log-parabola (LP), or a power law with super-exponential cutoff (PLSEC). Source positions and parameters are mapped onto a HEALPix-pixelated cube consisting of logarithmically spaced energy bins from 1 GeV to 1 TeV, and the final maps are expressed in units of ${\rm ph\ cm^{-2}\ s^{-1}\ sr^{-1}\ GeV^{-1}}$.
Each source is first assigned to the HEALPix pixel containing its sky position, yielding a ``delta'' point-source cube,
\begin{equation}
M_j(p)
=
\sum_{s\in p}
\frac{F_{s,j}}{\Omega_{\rm pix}\,\Delta E_j},
\qquad
\Delta E_j = E_{j+1}-E_j,
\end{equation}
where the sum runs over all catalog sources falling in pixel $p$, $F_{s,j}$ is the integrated flux of source $s$ in energy bin $j$, $\Omega_{\rm pix}$ is the solid angle of a HEALPix pixel, and $\Delta E_j$ is the bin width. Then, we produce a PSF-smoothed version of this cube by distributing each source's flux across neighboring pixels using the official Fermi-LAT point-spread function, modelled as a two-component King (Moffat) function\footnote{\url{https://fermi.gsfc.nasa.gov/ssc/data/analysis/documentation/Cicerone/Cicerone_LAT_IRFs/IRF_PSF.html}},
\begin{equation}
\mathrm{PSF}(\theta;\,E)
=
f_{\rm c}\,K(\theta;\,\sigma_{\rm c},\gamma_{\rm c})
+
(1-f_{\rm c})\,K(\theta;\,\sigma_{\rm t},\gamma_{\rm t}),
\qquad
K(\theta;\,\sigma,\gamma)
\propto
\left(1+\frac{\theta^2}{2\gamma\sigma^2}\right)^{-\gamma},
\end{equation}
where the core and tail widths are $\sigma_{\rm c,t}(E) = s_{\rm c,t}\,S(E)$, with $S(E)=\sqrt{[c_0(E/100\,\mathrm{MeV})^{\beta}]^2+c_1^2}$ the energy-dependent PSF scale factor. The parameters $s_{\rm c,t}$, $\gamma_{\rm c,t}$, $f_{\rm c}$, $c_0$, $c_1$ and $\beta$ are read directly from the Fermi CALDB response file \texttt{psf\_P8R3\_SOURCE\_V3\_PSF.fits}, averaged over the four PSF event types (PSF0--PSF3) and all inclination-angle bins. The kernel weights are evaluated in pixel space and normalized to unity,
\begin{equation}
M_j^{\rm PSF}(p)
=
\sum_{p'}
w_{pp'}\,M_j(p'),
\qquad
w_{pp'} \propto \mathrm{PSF}(\theta_{pp'};\,E_j^{\rm cen}),
\qquad
\sum_{p'} w_{pp'} = 1,
\end{equation}
where $\theta_{pp'}$ is the angular distance between pixels $p$ and $p'$ and $M_j(p')$ is the delta cube defined above. Normalizing the weights to unity guarantees exact flux conservation and avoids the negative ringing artifacts of harmonic-space smoothing.
Fig. \ref{fig:app_background} (\textit{right}) shows the different point-source contributions in the Galactic plane for the region $|b|<4^{\circ}$ constructing the template with four different approaches: (i.) using the best-fit spectral parameterization of the 4FGL-DR4 catalog, (ii.) assuming a PL, (iii.) a LP and (iv.) a PLSEC.
We find that a simple PL assumption systematically overestimates the point-source emission at high energies, whereas the PLSEC model leads to an underestimate relative to the catalog-based reconstruction. LP seems to agree well with the approach using the best-fit model. In the default analysis we use the PS-best contribution to model the point-source emission in the diffuse spectrum.

ii) \textit{The isotropic background ---} We employ the official \textit{Fermi}-LAT template (iso\_P8R3\_CLEAN\_V3\_v1)\footnote{\url{https://fermi.gsfc.nasa.gov/ssc/data/access/lat/BackgroundModels.html}}, which includes extragalactic emission and residual detector background. This contribution is not very important in the Galactic plane and it is the one with fewer uncertainties associated. 

iii) \textit{Fermi bubbles ---}
\label{app:fermi_bubbles}
The \textit{Fermi} bubbles are modeled as a separable component using a fixed spatial template mask $b(\hat n)$ and a tabulated energy spectrum $F(E)$ sourced from Ref. \cite{2014ApJ...793...64A}. This matches the public diffuse-background treatment and morphology/spectrum released with the LAT diffuse models~\cite{2017ApJ...840...43A} (see also the data products linked in the main text).
To construct the template, the binary mask is resampled to the HEALPix resolution of our analysis (matching the PASS8 and IEM maps) while preserving the published sky coverage.
The differential intensity at each analysis energy bin $E_j$ is determined via linear interpolation of the tabulated values, with the intensity set to zero outside the reported range. 
Within our Galactic-plane regions of interest (ROIs), the bubble contribution is calculated as the weighted mean of the masking intensity, effectively assuming a spatially uniform spectral energy distribution (SED).
Although the predicted bubble flux is subdominant to the hadronic diffuse emission in the disk, this component is included to ensure a complete and standard decomposition of the $\gamma$-ray sky.

iv) \textit{Inverse-Compton (IC) ---}
We model the IC component following Ref.~\cite{DeLaTorreLuque:2025zsv}, where the steady-state Galactic electron/positron distribution is computed with \textsc{DRAGON2} and tuned to local CR electron and positron measurements. A key ingredient is the adoption of a 3D source distribution tracing the Milky Way spiral arms, which becomes relevant once radiative losses (IC and synchrotron) shape the high-energy lepton spectrum; while both 2D and spiral-arm setups reproduce local data, the spiral-arm case implies a harder injection and yields a larger average IC emission in the Galactic plane. Above a few GeV, bremsstrahlung is negligible, and the IC component remains subdominant in the plane at the \(\sim 10\%\) level for \(E \gtrsim 5\)~GeV.

v) \textit{Bremsstrahlung  ---}
For bremsstrahlung, we follow Ref.~\cite{DeLaTorreLuque:2025zsv}, adopting the same CR lepton modeling used for the IC component. This contribution is relevant only at low energies and becomes negligible above a few GeV.

vi) \textit{Unresolved sources} --- We adopt the model used in Refs.~\cite{Luque:2022buq, DeLaTorreLuque:2025zsv}, and derived in Ref.~\cite{Steppa2020} for HESS data.

Besides the non-hadronic components, we add to Fig.~\ref{fig:app_background} the expected spectrum from the $\pi^0$ component derived from a CR spectrum (obtained with DRAGON~\cite{Evoli_2017}) following the PL that reproduces the proton data below 300 GeV (green line). This component is not used in our analysis and is included solely for illustration, showing that the convolution of a single PL for the proton spectrum and the gamma-ray production cross sections leads to a PL gamma-ray spectrum (computed using Hermes~\cite{Dundovic_2021}). The sum of this $\pi^0$ component and the non-hadronic emission is shown as a gray line. 

\section{SED Plots}
\label{app:plots}

\begin{figure}[t]
    \centering
    \includegraphics[width=0.99\linewidth]{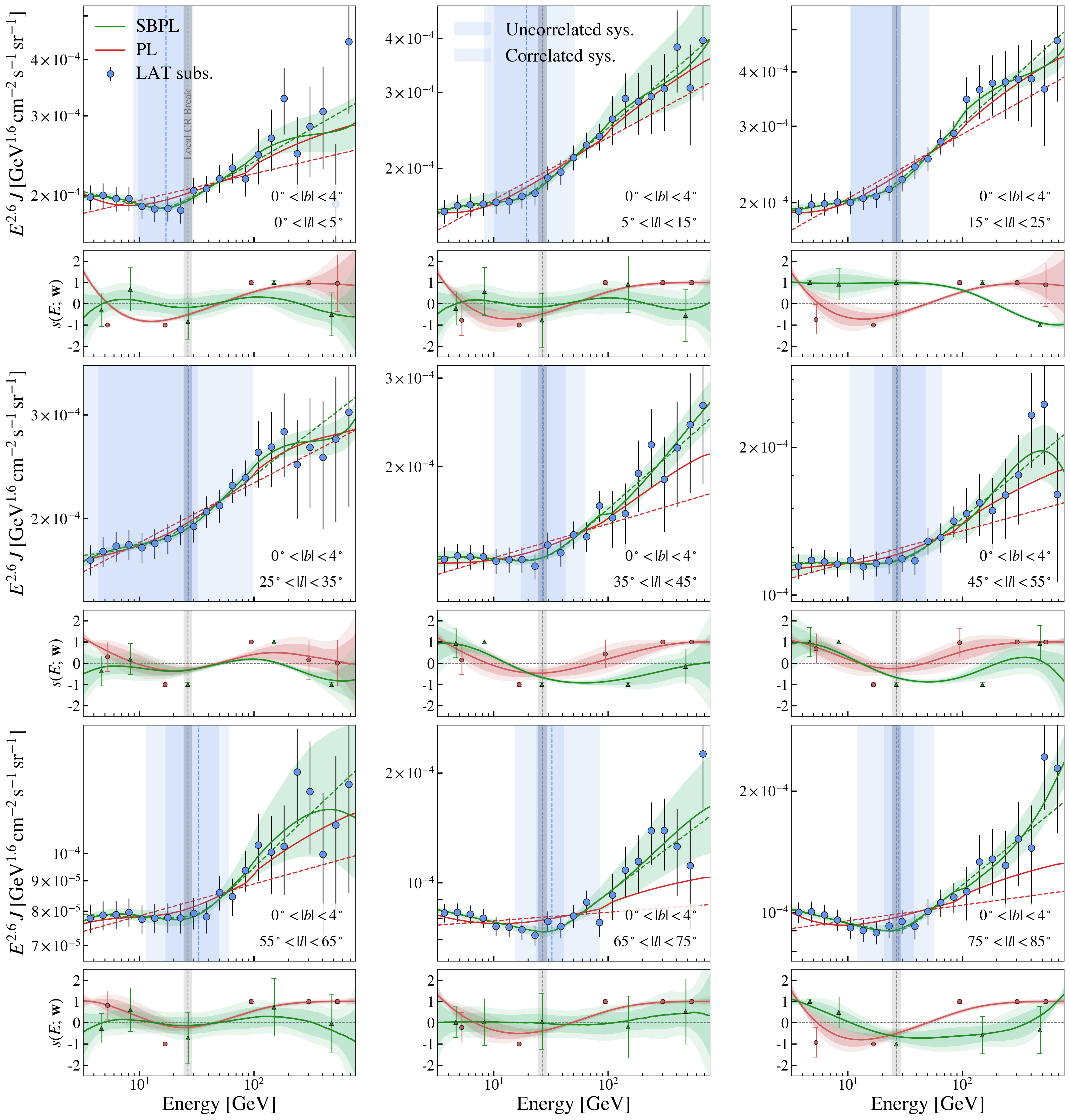}
    \caption{Window-by-window SEDs for $|b|<4^\circ$ and $0^\circ<|l|<85^\circ$ (9 windows) for \textit{Fermi}-LAT data. In each cell, the top panel shows the measured spectrum together with the best-fit PL and SBPL models; the grey vertical band marks the reference ``Local CR Break'' energy range and the shaded blue region indicates the fitted $E_{\rm break}$ interval. The bottom panel shows the deformation function $s(E;w)$ as a function of energy and the corresponding nuisance coefficients $w_j$. 
}
\label{fig:app_pass8_nuisance_plots}
\end{figure}

In Fig.~\ref{fig:app_pass8_nuisance_plots} we present the nine Galactic-longitude windows used in the analysis described in the main text, showing the spectral energy distribution (SED) together with the distortion function \(s(E,w)\), which models effective-area systematic uncertainties and their correlations with energy. The figures clearly show evidence for a spectral break in the diffuse emission across all longitude windows, remarkably consistent with the break observed in local CR data (grey band). We observe that the break significance decreases at intermediate longitudes and becomes more pronounced at larger longitudes, where the impact of diffuse background emission is reduced. Under the null hypothesis (no break), the nuisance distortion tends to mimic a break-like feature---positive at low energies, negative at intermediate energies, and positive again at high energies---whereas in the SBPL model the distortion remains close to zero.

\section{Systematic uncertainties as nuisance parameters}
\label{app:nuisance_systematics}

To test the robustness of the spectral break against potential systematic bias in the \textit{Fermi}-LAT measurement induced by effective-area uncertainties, we model the instrumental uncertainty as  a sequence of unknown correction factors. They correspond to nuisance parameters described as a smooth deformation of the predicted spectrum. This follows the statistical approach of previous
\textit{Fermi}-LAT analyses~\citep{2017PhRvD..95h2007A}, where
energy-dependent systematics are encoded through nuisance functions and
profiled together with the physical spectral parameters.
For a model $M$ (PL or SBPL), the prediction in energy bin $i$ is $1+s(E_i; \textbf{w})S(E_i)$ and the $\chi^2$ function for the spectral fit is given by
\begin{equation}
\chi^2_{k,M}(\theta,\mathbf w)
=
\sum_{i}
\left[
\frac{
y_{ik}
-
\left(1+S_{ik}s(E_i;\mathbf w)\right)
\mu_{ik}^{M}(\theta)
}{
\sigma_{ik}
}
\right]^2
+
\chi^2_{\rm prior}(\mathbf w).
\label{eq:chi2_with_nuisance}
\end{equation}
where $\theta$ are the free parameters: $\theta_{\rm PL}=\{K, \gamma\}$ and $\theta_{\rm SBPL}=\{K, \gamma,\Delta\gamma, E_{\rm break}\}$, while $\mu_{ik}^{M}(\theta)$ is the nominal physical model prediction, $\sigma_{ik}^2$ is the quadratic sum of the statistical and acceptance uncertainties and $S_{ik}$ is the fractional systematic error (effective area error, in this case). Then, $\mathbf w=\{w_1,...,w_{N_{\rm nuis}}\}$ denotes the vector of nuisance amplitudes 
per longitude window $\textbf{w}_k$ and $s(E;\mathbf w)$ is a dimensionless distortion function, which smoothly connects the nuisance parameters as:
$x=\log_{10}E$
\begin{equation}
s(x;\mathbf w)
=
\sum_{j=1}^{N_{\rm nuis}} w_j B_j(x).
\label{eq:nuisance_basis}
\end{equation}
where $B_j(x)$ are the standard B-spline basis functions of degree $d=3$ on an open-uniform knot vector, with $N_{\rm nuis}$ control coefficients placed at Greville abscissae, acting as spline weights enforcing smooth energy-correlated deformations.
The number of nuisance parameters $N_{\rm nuis}$ cannot be inferred from physical principles, our choice of $N_{\rm nuis}=6$ places one nuisance node per $0.5$ decades of $\log_{10}(E)$, following the bracketing method recommended by the \textit{Fermi}-LAT collaboration \footnote{\url{https://fermi.gsfc.nasa.gov/ssc/data/analysis/scitools/Aeff_Systematics.html}}. This sampling is fine enough to faithfully track the smooth energy dependence of the LAT effective-area uncertainty.
The second term in Eq.~\ref{eq:chi2_with_nuisance} $\chi^2(\textbf{w})$ penalizes nuisance excursions beyond the allowed envelope, the prior follows a box-Gaussian distribution inside the quoted systematic envelope, with a Gaussian tail penalization,
\begin{equation}
\chi^2_{\rm prior}(w_j)
=
\begin{cases}
0, & |w_j|\le s_w,\\[6pt]
\dfrac{(|w_j|-s_w)^2}{\tau^2}, & |w_j|>s_w.
\end{cases}
\label{eq:box_gauss_penalty}
\end{equation}
Each longitude window $k$ has its own nuisance vector $\textbf{w}_k$, so nuisance amplitudes are profiled out and the best-fit parameters are obtained with uncertainties derived from one-dimensional profile-likelihood scans.
For the \textit{global} fit, where break parameters are shared across all windows, we have the following $\chi^2$
\begin{equation}
\chi^2_{{\rm glob},M} = \min_{\{K_k,\gamma_{1,k}\},\{\mathbf{w}_k\},E_{\rm br}^{\rm glob},\Delta\gamma^{\rm glob}} \sum_{k=1}^{N_{\rm win}} \chi^2_{k,M}(\theta_k,\mathbf{w}_k),
\label{eq:chi2_global}
\end{equation}

\noindent
\textbf{Pseudo-experiments under the null hypothesis --- }
Since nuisance profiling and bounded priors can violate simple asymptotic assumptions, we calibrate the significance with Monte Carlo pseudo-experiments generated under the profiled PL null hypothesis. For each pseudo-experiment, in window $k$ and energy bin $i$ we draw
\begin{equation}
y_{ik}^{\rm pseudo}
=\left[1+S_{ik}\,s(E_i;\mathbf w^{\ast})\right]\mu_{ik}^{\rm PL}(\boldsymbol{\theta}^{\ast})
+\epsilon_{ik},
\label{eq:pseudo_data}
\end{equation}
where \(\epsilon_{ik}\sim\mathcal N(0,\sigma_{ik}^2)\) models the statistical scatter, with \(\sigma_{ik}\) the per-bin uncertainty used in the \(\chi^2\) fits. The PL parameters \(\boldsymbol{\theta}^{\ast}\) are drawn from the best-fit null model, accounting for the corresponding parameter covariance when available.
Each pseudo-experiment is analyzed with exactly the same fit configuration as the data: the PL null and the alternative model are refitted. For each pseudo-dataset \(n\), we compute the $\Delta\chi^2_{{\rm pseudo},n}$ from the corresponding profiled fit statistics (cf. Eq. \ref{eq:chi2_with_nuisance}) and compare it with the observed \(\Delta\chi^2_{\rm obs}\). The MC \(p\)-value is defined as the fraction of null pseudo-experiments that yield an improvement at least as large as observed.
\begin{equation}
p_{\rm MC}=\frac{1}{N_{\rm MC}}\sum_{n=1}^{N_{\rm MC}}\mathbf{1}\!\left(\Delta\chi^2_{{\rm pseudo},n}\ge \Delta\chi^2_{\rm obs}\right),
\label{eq:pvalue_mc}
\end{equation}
with finite-sample regularization $p_{\rm MC}\ge 1/(N_{\rm MC}+1)$.
The corresponding one-sided Gaussian-equivalent significance is
$Z_{\rm MC}=\Phi^{-1}(1-p_{\rm MC})$
where $\Phi$ is the standard normal CDF. We also quote a Wilks-based reference,
$p_{\rm Wilks}=P\!\left(\chi^2_\nu\ge \Delta\chi^2_{\rm obs}\right)$
typically with $\nu=2$ for $(\Delta\gamma,E_{\rm break})$. Because nuisance profiling and box-constrained priors make the setting non-regular, MC calibration is taken as the primary significance measure.

\section{Robustness tests}
\begin{figure}
    \centering
    \includegraphics[width=\linewidth]{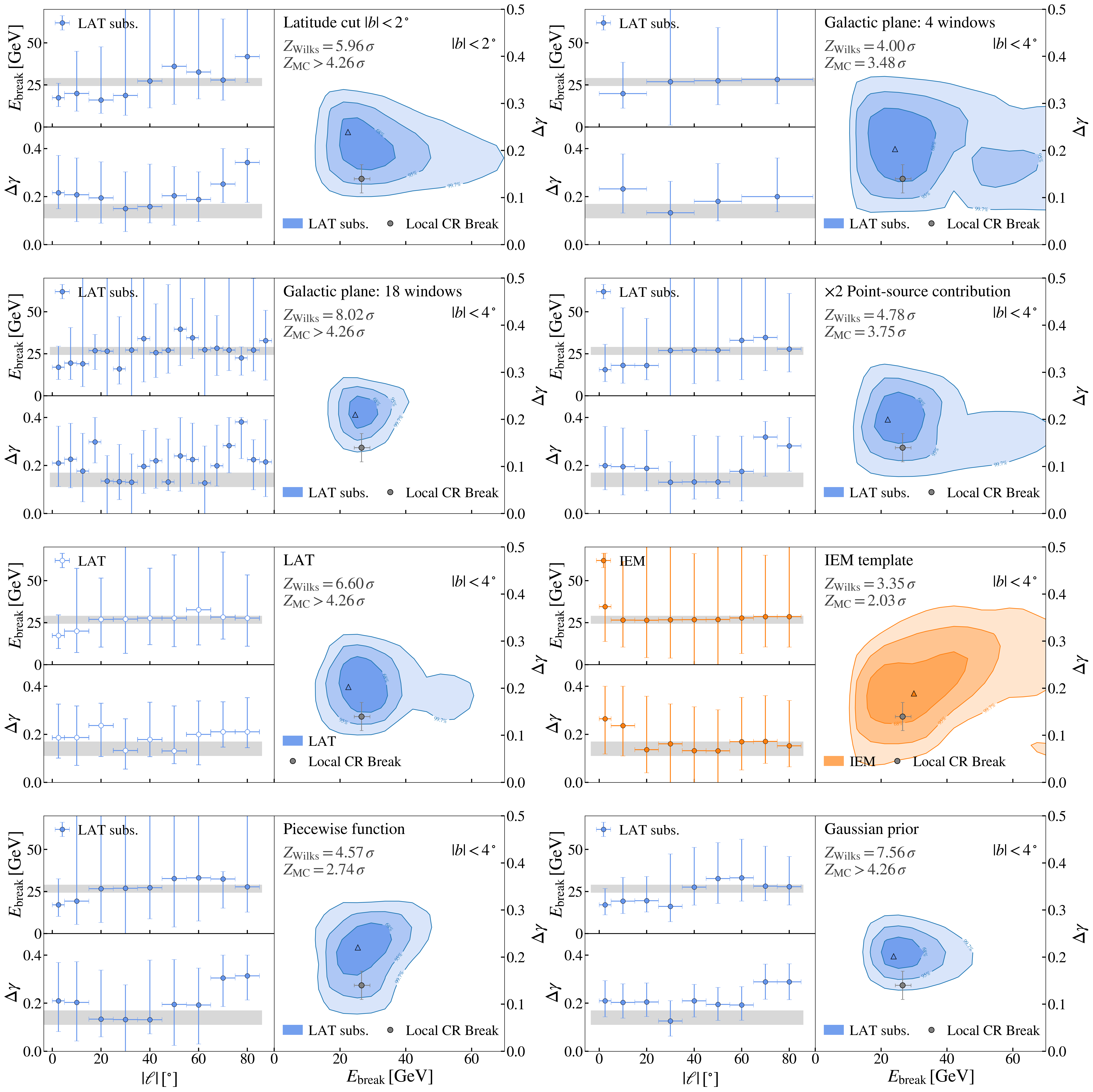}

\caption{Robustness tests:
\textit{Left}: Best-fit break energy $E_{\mathrm{break}}$ (top) and spectral-index change $\Delta\gamma$ (bottom) as a function of Galactic longitude $|\ell|$, inferred from \textit{Fermi}-LAT data (blue) or the IEM template (orange, row~3 right). Grey bands indicate the local cosmic-ray break values.
\textit{Right}: Joint confidence contours ($1\sigma$, $2\sigma$, $3\sigma$) in the $(E_{\mathrm{break}},\,\Delta\gamma)$ plane from a global fit assuming a common break across the corresponding longitude windows; the triangle marks the best-fit point and the grey marker the local cosmic-ray measurement.
Each row corresponds to a different systematic variation:
(\textit{row~1, left}) narrower latitude selection;
(\textit{row~1, right}) 4 Galactic-plane longitude windows;
(\textit{row~2, left}) 18 longitude windows;
(\textit{row~2, right}) doubled point-source contribution;
(\textit{row~3, left}) \textit{Fermi}-LAT unsubtracted data;
(\textit{row~3, right}) IEM diffuse-emission template;
(\textit{row~4, left}) non-smooth piecewise spectral function;
(\textit{row~4, right}) Gaussian priors on nuisance parameters.
The significance values $Z_{\mathrm{Wilks}}$ and $Z_{\mathrm{MC}}$ quoted in each panel refer to the detection significance of the spectral break under Wilks' theorem and Monte Carlo calibration, respectively.%
}
\label{fig:tests}
\end{figure}

We present the tests described in the main text for the robustness of our results 
under variations in foreground components and analysis choices, all of them 
summarised in Fig.~\ref{fig:tests}.
\\

\noindent
\textit{i) Analysis regions ---}
We tested the same pipeline by varying the latitude cuts and longitude window sizes: imposing a latitude cut $|b|<2^\circ$, dividing the Galactic plane in 4 windows and in 18 longitude windows. The evidence for a spectral break persists even after accounting for correlated systematics. Although narrower windows lead to larger uncertainties, they allow for a better determination of the break position when the break parameters are imposed to be common along all the longitude windows.
\\

\noindent
\textit{ii) Foreground components ---}
We assess the impact of foreground subtraction both individually and in combination, finding no significant variation in the spectral break parameters. As shown in Fig.~\ref{fig:app_background}, the diffuse emission in the Galactic plane is dominated by hadronic emission, with all remaining components at least an order of magnitude lower. Nevertheless, since the point-source contribution modelled from the 4FGL-DR4 catalogue could in principle be responsible for the observed break — as also suggested by local source injection scenarios for the 300 GV feature — we explicitly test this by adopting a point-source normalisation 100\% higher than the measured value. The recovered break parameters are fully consistent with those reported in the main text. As a further check, we perform the fit directly on the unsubtracted \textit{Fermi}-LAT data, again finding consistent results.
\\

\noindent
\textit{iii) IEM consistency-check ---}
The official \textit{Fermi}-LAT Galactic Interstellar Emission Model (IEM) provides a complementary characterisation of the diffuse emission, encoding hadronic, IC, and bremsstrahlung contributions through the GALPROP framework tuned to LAT data~\cite{Fermi-LAT:2016zaq}. 
As a consistency check, processing the IEM template through an identical analysis pipeline yields break parameters fully consistent with those derived from the \textit{Fermi}-LAT data directly, confirming that our foreground prescription does not artificially generate or suppress the feature.
\\

\noindent
\textit{iv) Nuisance configurations ---} We tested variations on the assumption of the nuisance configurations used to model the effect of effective-area correlated systematics.
\begin{enumerate}[itemsep=0pt, nosep]
    \item Smoothness: instead of a smooth B-spline parameterization we can use a piecewise-linear basis in
    $\log_{10}E$.  In this alternative parameterisation, $N_{\rm nuis}$ reference energies $x_j$ are placed uniformly across the fitted energy range and the nuisance parameters have the direct interpretation $w_j=s(x_j)$ with a linear interpolation. 
    \item Priors: instead of box-uniform priors we could consider Gaussian nuisance priors,
    \begin{equation}
    \chi^2_{\rm prior}(\mathbf w)
    =
    \sum_j w_j^2,
    \label{eq:gaussian_nuisance_prior}
    \end{equation}
    corresponding to independent unit-variance Gaussian priors centred at zero.
    This is closer to the nuisance prescription used in a previous
    \textit{Fermi}-LAT systematic study~\citep{2017PhRvD..95h2007A}. The two
    approaches have the same likelihood structure; they differ only in how
    strongly non-zero deformations are penalized.
\end{enumerate}

\section{Tables}

\begin{table}[h!]
\centering
\label{tab:BPL_summary}
\renewcommand{\arraystretch}{1.1} 
\begin{tabular*}{\columnwidth}{@{\extracolsep{\fill}}l|c|ccccc}
\hline\hline
\textbf{SBPL} & $l$ & $\Delta\chi^2$ & $E_{\rm break}$ [GeV] & $\Delta\gamma$ & $Z_{\rm W}$ [$\sigma$] & $Z_{\rm MC}$ [$\sigma$]\\
\hline
 & 0-5   & 2.96  & $17.05_{-7.35}^{+12.48}$  & $0.21_{-0.10}^{+0.15}$ & 0.75 & 1.10 \\
              & 5-15  & 2.81  & $19.28_{-11.09}^{+31.76}$ & $0.20_{-0.12}^{+0.15}$ & 0.69 & 1.09 \\
& 15-25 & 4.10  & $26.75_{-16.41}^{+24.14}$ & $0.25_{-0.14}^{+0.09}$ & 1.13 & 1.50 \\
       & 25-35 & 1.16  & $26.97_{-22.67}^{+71.44}$ & $0.13_{-0.11}^{+0.12}$ & -0.15 & 0.93 \\
                    & 35-45 & 4.21  & $27.35_{-17.21}^{+35.81}$ & $0.18_{-0.09}^{+0.16}$ & 1.17 & 1.55 \\
                    & 45-55 & 2.90  & $27.31_{-16.94}^{+38.81}$ & $0.13_{-0.06}^{+0.19}$ & 0.72 & 1.19 \\
                    & 55-65 & 2.63  & $33.16_{-21.69}^{+27.87}$ & $0.19_{-0.10}^{+0.14}$ & 0.62 & 1.02 \\
                    & 65-75 & 9.51  & $32.42_{-17.19}^{+52.60}$ & $0.31_{-0.11}^{+0.05}$ & 2.38 & 2.65 \\
                    & 75-85 & 9.72  & $27.67_{-15.64}^{+29.44}$ & $0.24_{-0.08}^{+0.19}$ & 2.42 & 2.65 \\
\hline
Global              & 0-90  & 39.78 & $24.08_{-1.63}^{+9.06}$   & $0.20_{-0.02}^{+0.05}$ & 5.86 & $4.44$ \\
Fixed               & 0-90  & 35.49 & 26.52                     & 0.140                    & ---  & $4.28$ \\
\hline\hline
\end{tabular*}
\caption{Summary of fit quality and spectral break parameters across longitudes.
For each model case, we report the $\Delta\chi^2$, the best-fit break energy
$ E_{\rm break}$ (GeV), and spectral change
$\Delta\gamma$, and significance in terms of Wilks' theorem ($Z_{\rm Wilks}$) and MC pseudo-experiments under the null hypothesis ($Z_{\rm MC}$) for \textit{Fermi}-LAT data.}
\end{table}


\bibliographystyle{apsrev4-1}
\bibliography{CRbiblio}